\documentstyle[12pt]{article}
\topmargin--0cm
\oddsidemargin--1mm
\begin{document}

\begin{center}
{\LARGE Spaces with Torsion from Embedding,\\ and the Special Role of
Autoparallel Trajectories}
\vskip 0.5cm
{Hagen KLEINERT\footnote{
Email: kleinert@physik.fu-berlin.de; shabanov@physik.fu-berlin.de; URL:
http://www.physik.fu-berlin.de/\~{}kleinert. Phone/Fax:
 0049/30/8383034 }
\  and\  Sergei V. SHABANOV}\footnote{\noindent
on leave from Laboratory of Theoretical
Physics, JINR, Dubna, Russia; a DFG fellow.}
\vskip 0.5cm {\em Institute for Theoretical Physics, FU-Berlin,
Arnimallee 14, D-14195, Berlin, Germany}
\end{center}

\begin{abstract}
As a contribution to the ongoing discussion
of trajectories of spinless particles
in spaces with torsion we
show that the geometry of such spaces can be
induced by embedding their curves in a euclidean space
without torsion.
Technically speaking, we define
the tangent (velocity) space of the embedded space imposing non-holonomic
constraints upon the  tangent space of the embedding space.
Parallel transport in the embedded space
is determined as an induced
parallel transport on the surface of constraints.
Gauss' principle
of least constraint is used to show that autoparallels
realize a constrained motion that has
a {\em minimal} deviation from the free, unconstrained motion,
this being a mathematical expression of the principle of inertia.
In contrast, geodesics play no special role in the constrained
dynamics, making them less likely candidates for particle trajectories.
\end{abstract}

\noindent
{\bf 1}.  On an affine manifold equipped with a metric,
there exist two preferred connections compatible
with the metric \cite{schouten}.
One is the Riemann connection
defined only by the metric. In
a coordinate basis in the tangent space of the manifold,
the coefficients of this connection are Christoffel symbols
\begin{equation}
\bar{\Gamma}_{\mu\nu\kappa} =g_{ \kappa\lambda}
\bar{\Gamma}_ {\nu\kappa}{}^ \lambda =
{\textstyle{\frac 12}}\left(g_{\mu\nu,\kappa} +
g_{\mu\kappa,\nu} - g_{\nu\kappa,\mu}\right)\ .
\label{cristoffel}
\end{equation}
Here $g_{\mu\nu}$ are components of the metric tensor, and
indices after a comma stand for the corresponding derivatives,
$T_{\mu\nu...,\lambda \kappa...} = \partial_\lambda\partial_ \kappa
\, {\cdots}\, T_{\mu\nu...}$.
By definition, the covariant derivative
formed with this
Riemann connection
satisfies the {\em metricity condition\/}
\begin{equation}
\bar D_\mu g_{\nu\kappa} = \partial_\mu g_{\nu\kappa} -
\bar \Gamma_{\nu\mu}{}^\lambda g_{\lambda\kappa} -
\bar\Gamma_{\kappa\mu}{}^\lambda g_{\nu\lambda} =0\ .
\label{dg0}
\end{equation}
Apart from $\bar{\Gamma}_{\mu\nu}{}^{\lambda}$, there exists
also a Cartan connection
$\Gamma_{\mu\nu}{}^\kappa$.
It satisfies the same
compatibility condition with the metric:
\begin{equation}
D_\mu g_{\nu\kappa} = \partial_\mu g_{\nu\kappa} -
\Gamma_{\nu\mu}{}^\lambda g_{\lambda\kappa} -
\Gamma_{\kappa\mu}{}^\lambda g_{\nu\lambda} =0\ .
\label{dg}
\end{equation}
It can always be represented in the form \cite{schouten}
\begin{equation}
\Gamma_{\mu\nu\kappa} = \bar{\Gamma}_{\mu\nu\kappa} +
K_{\mu\nu\kappa}\ ,
\label{gamma}
\end{equation}
where $K_{\mu \nu \kappa}$ is any antisymmetric tensor in $ \nu \kappa$,
called the contorsion tensor
\cite{schouten}
\begin{equation}
K_{\mu\nu\kappa} =
S_{\mu\nu\kappa} - S_{\nu\kappa\mu} + S_{\kappa\mu\nu}\ ,
\label{K}
\end{equation}
where
 $S_{\mu\nu\kappa} =
g_{ \kappa\lambda}S_{\mu \nu}{}^ \lambda$ and
$S_{\mu \nu}{}^ \lambda$ is the torsion tensor
\begin{equation}
S_{\mu \nu}{}^ \lambda = \frac 12\left(
\Gamma_{\mu \nu}{}^ \lambda -\Gamma_{ \nu\mu}{}^ \lambda \right)\ .
\label{torsion}
\end{equation}

Each connection compatible with the metric on the manifold defines
a curve which parallel-transports
its tangent vector along itself. Let $v^\mu$ stand for the tangent vector
and $\dot{v}^\mu$ for its derivative with respect to the
affine parameter, the proper time
on the curve. Then the equation for the curve which parallel-transports
its tangent vector along itself with respect to the Cartan connection
reads
\begin{equation}
\frac{Dv^\mu}{dt} \equiv
\dot{v}^\mu + \Gamma_{\lambda\nu}{}^ \mu v^\lambda v^ \nu = 0\ .
\label{auto}
\end{equation}
It describes {\em autoparallels\/}, the {\em straightest\/} curves
in Riemann-Cartan space.
The reason for this name will be explained later in Section 5.
The same equation with the
Christoffel symbols
 describes {\em geodesics\/},
the {\em shortest\/} curves with respect to the metric $g_{\mu \nu}$.
For the Cartan connection (\ref{gamma}),
the deviation from the geodesics is caused by a torsion force
in (\ref{auto}) coming from the
symmetrical part of the contorsion tensor
$K_{\{\mu\nu\}\lambda}v^\mu v^ \nu =-2 S_{\mu \lambda\nu} v^\mu v^ \nu $.

Apart from extremizing
a length between two fixed endpoints,
geodesics in a  Riemannian space
can be obtained by embedding the Riemannian space
in a euclidean space of a higher dimension.
This is done by imposing certain constraints
on the Cartesian coordinates spanning
the euclidean space.
The points on the constraint surface
constitute the embedded Riemannian space.
Straight lines in the euclidean space, which are
geodesic and autoparallel and also determine a free motion in that space,
become geodesics when the motion is restricted to
the constraint surface.
The restriction of the free motion to the constraint surface
is done in a conventional way, i.e., by adding the equations of
constraints
to the equations of motion. When the constraint force is removed,
geodesic trajectories  turns into straight lines in the embedding
space.

For dynamics in Riemann-Cartan spaces, no such embedding
is known. The purpose of this letter is to fill this gap.
The new embedding will have the property that autoparallels
are realized as trajectories of a constrained free motion.

Key observation of our theory is the fact that
in order to define a geometry (metric and the law
of parallel-transport) by embedding it is
not necessary to constrain the positions in a euclidean space.
One may impose constraints only on the
tangent (or velocity) space,
thereby defining a physical tangent space as a subspace embedded
in a bigger velocity space. This is
sufficient to define all curves in the embedded space as a special
subset of all curves in the embedding space because each curve
is specified by its tangent vector. The metric in the embedded space
is naturally {\em induced\/}
by restricting the scalar product in the bigger (embedding) tangent
space to the constraint surface. The {\em induced connection\/} is
uniquely determined by the compatibility condition of the
embedding of the tangent space
with the parallel-transport law in the embedding space.
This means the following: Take a curve in the original
space connecting points 1 and 2. This curve is then embedded
into a bigger euclidean space by specifying the tangent vector
of the image curve. A vector from the tangent space at point 1 is
parallel-transported along the curve to point 2, and then it is
embedded in the bigger space. We require that the resulting
vector must be the same as the one obtained in the opposite way:
The vector at point 1 is first embedded into a bigger
euclidean space and then parallel-transported along the image
of the curve connecting points 1 and 2. This compatibility
condition ensures that
the connection in the original space is uniquely determined
by the embedding law.

Constraints imposed on the tangent space
can be non-holonomic, and this is the source of torsion.
The notion of "holonomic"
and "non-holonomic" constraints is the same as in classical mechanics.
For a mechanical system, generalized velocities are
elements of the tangent space of its configuration
space. Let the motion be subject to constraints linear in velocities.
According to the
Hertz classification \cite{arnold}, constraints are said
to be holonomic
if they are integrable (i.e., equivalent to some constraints
on the configuration space only), and non-holonomic if
they are non-integrable. Sometimes dynamical systems with
non-holonomic constraints are simply called non-holonomic
systems. It is important to realize that the motion of
non-holonomic systems does not occur on any submanifold
of the configuration space, nonetheless it is described
by a less number of parameters than the
corresponding unconstrained motion.

Upon an embedding via non-holonomic
constraints on the tangent space,
any curve that parallel-transports its tangent vector
along itself has an image with same property. Therefore
straight lines, which are autoparallels and geodesics with respect
to a trivial euclidean connection, are natural images
of {\em autoparallels\/} in the embedded space.
Using Gauss' principle of least constraint
we show that autoparallels describe a constrained motion such that the
acceleration (or the force) induced by the constraints has
a {\em least} deviation from the acceleration of the corresponding
unconstrained motion, while geodesics play no special role in the
constrained dynamics.
This comprises the main result of our paper.

We remark that torsion has  been included into
general relativity. The corresponding theory is known as
the Cartan-Einstein theory \cite{thorn}.
 From the point of view of the general coordinate invariance,
both geodesics and autoparallels are equally good candidates
for trajectories of  spinless test particles \cite{auto}.
The conventional way of obtaining the law of interaction
between matter and spacetime connection is to apply the
gauge principle and minimal coupling \cite{kibble}.
In such an approach, scalar fields
are decoupled from torsion, thus leading to geodesics
as true trajectories of spinless point particles. This, however,
does not imply that the autoparallels are incompatible with
the gauge principle. Equation (\ref{auto}) is just as gauge
covariant as the geodesic equation $\bar{D}v^\mu/dt =0$.
But autoparallels do not seem to be compatible with the
{\em minimal} gauge coupling principle, when all derivatives
in a Lagrangian of a {\em free} theory are replaced by
the covariant derivatives in order to obtain the interacting
theory.

The embedding has been a powerful tool in studying geometrical forces
in non-euclidean configuration spaces.
Our approach based on the constrained dynamics
might be useful in constructing
possible field models of interaction between matter and
a generic spacetime connection where
autoparallels are true trajectories of spinless particles.
Up to now neither experiments nor theoretical
principles forbid such theories.

\vskip 0.3cm
\noindent
{\bf 2}. Let $[x]$ be a Euclidean space and $x^i, i=1,2,...,M=\dim [x]$,
be a set of coordinates. Let $[q]$ be a space of a smaller dimension,
$N=\dim [q]$, spanned by local coordinates $q^\mu,\ \mu =1,2,...,N$.
Tangent spaces of $[x]$ and $[q]$ are denoted as $T[x]$ and $T[q]$,
respectively. Consider a curve $q^\mu(t)$ in $[q]$ and tangent spaces
at each point of the curve, $T[q(t)]$. We define a curve $x^i(t)$ in
the embedding space $[x]$ by specifying its tangent vector
\begin{equation}
v^i =\varepsilon^i{}_\mu(q)v^\mu\ , \ \ \ v^\mu = \dot{q}^\mu(t)\ ,
\ \ \ v^i = \dot{x}^i(t)\ ,
\label{tv}
\end{equation}
where coefficients $\varepsilon^i{}_\mu(q)$ are some functions on $[q]$.
 From (\ref{tv}) follows
\begin{equation}
x^i(t) = x^i(0) + \int\limits_{0}^{t}dq^\mu \varepsilon^i{}_\mu(q)\ .
\label{xcurve}
\end{equation}
For any curve $q^\mu(t)$ in $[q]$, equation (\ref{xcurve})
determines a curve in the space $[x]$ up to a global translation on
a vector $x^i(0)$. Thus it determines an embedding of the space
of all paths in $[q]$ into the space of all paths in $[x]$. Assuming
equation (\ref{tv}) to hold for all curves in the space $[q]$ passing
through some point $q^\mu$, we
specify the embedding of $T[q]$ at this point into the Euclidean
space $T[x]$. Note that
tangent spaces at all points
of $[x]$ are the same and coincide with the space $[x]$
because the parallel
transport is trivial in the space $[x]$ (the connection $\Gamma_{ij}{}^k$
vanishes identically). So, the shift of the image curve $x^i(t)$ on
a constant vector in (\ref{tv}) is irrelevant for the embedding
of the tangent space.
The embedding of the path space or the tangent space
includes the case of the pointwise embedding of the space $[q]$ itself into
$[x]$, but it appears to be more general as we shall see.

Consider
two sets of tangent spaces $T[q(t)]$ and $T[x(t)]$
at points of the curve $q^\mu(t)$ and of its image $x^i(t)$
defined by (\ref{xcurve}). We embed the space $T[q(t)]$
in $T[x(t)]$ so that for any element $T^\mu$ of
$T[q(t)]$, its image in $T[x(t)]$ is
\begin{equation}
T^i = \varepsilon^i{}_\mu(q(t)) T^\mu\ .
\label{emb}
\end{equation}
We perform a parallel transport
of the image vector $T^i$ along the curve
$x^i(t)$. Since the connection
in the space $[x]$ is zero, an infinitesimal change of $T^i$ is
determined by the total derivative with respect to the affine
parameter
\begin{equation}
\frac{DT^i}{dt} = v^jD_jT^i = \frac{dT^i}{dt}\ .
\label{dti}
\end{equation}
Similarly,  an infinitesimal change of the vector
$T^\mu$ under the parallel transport along the curve $q^\mu(t)$
is specified by the covariant derivative
\begin{equation}
\frac{DT^\mu}{dt} = v^\nu D_\nu T^\mu =
v^\nu\left(\partial_\nu T^\mu + \Gamma_{ \kappa\nu}{}^\mu  T^\kappa\right)\
\label{dt}
\end{equation}
with $\Gamma_{\mu\nu}{}^\lambda$ being the connection on the space $[q]$.
Now we come to the crucial condition for our embedding procedure:
We require that the vector $DT^i/dt$ must be the image
of the vector $DT^\mu/dt$, that is,
\begin{equation}
\frac{DT^i}{dt} = \varepsilon^i{}_\mu\,\frac{DT^\mu}{dt}\ .
\label{embcon}
\end{equation}
Equation (\ref{embcon}) has a transparent geometrical meaning.
The parallel transport of the vector $T^\mu$ along any curve
$q^\mu(t)$ and the subsequent embedding of the resulting vector
into the bigger space $T[x]$ via the relation (\ref{emb})
 gives an element of $T[x]$.
This element must coincide with the one obtained in the opposite
way in which the vector $T^\mu$ is first embedded  and then
parallel-transported along the image
$x^i(t)$ of the curve $q^\mu(t)$. This implies that the embedding
of the tangent space $T[q]$ at any point of $[q]$ is compatible
with the parallel transport on $[q]$.
The compatibility condition (\ref{embcon})
uniquely determines the connection coefficients
$\Gamma_{ \nu\kappa}{}^\mu$ via the embedding
coefficients $\varepsilon^i{}_\mu$.

Before we proceed to prove this statement, let us introduce some
useful notations. For any two vectors from $T[x]$, one can introduce
a scalar product associated with the Cartesian metric on $[x]$
\begin{equation}
(\tilde{T},{T}) = \delta_{ij}\tilde{T}^i{T}^j\ .
\label{sp}
\end{equation}
If the vectors $T^i$ and $\tilde{T}^i$
are the images of $T^\mu$ and $\tilde{T}^\mu$, respectively, then
the embedding coefficients $\varepsilon^i{}_\mu$
determine an {\em induced metric\/} on $[q]$
\begin{equation}
(\tilde{T},{T}) = g_{\mu\nu} \tilde{T}^\nu{T}^\mu\ ,\ \ \ g_{\mu\nu} =
(\varepsilon_\mu,\varepsilon_\nu)\ .
\label{met}
\end{equation}
It is useful to introduce the quantity
\begin{equation}
\varepsilon^{i\mu} = \varepsilon^i{}_{\nu} g^{\nu\mu}\ ,
\label{covec}
\end{equation}
where $g^{\mu\lambda}g_{\lambda\nu} = \delta^\mu_\nu$. From (\ref{met})
follows that
\begin{equation}
(\varepsilon^\mu,\varepsilon^\nu) = g^{\mu\nu}\ ,
\ \ \ (\varepsilon^\mu,\varepsilon_\nu)=
\delta_\nu^\mu\ .
\label{invmet}
\end{equation}
Assuming that
the metrics $g_{\mu\nu}$ and $\delta_{ij}$ are used to lower or raise
indices of tensors on $[q]$ and $[x]$, respectively, the embedding
condition (\ref{embcon}) can be written in a more general form
\begin{equation}
\frac{dT^{ij\cdots}_{kn\cdots}}{dt} =
\frac{d}{dt}\left(
\varepsilon^i{}_\mu\varepsilon^j{}_\nu\cdots
\varepsilon_k{}^\lambda\varepsilon_n{}^\beta\cdots
T^{\mu\nu\cdots}_{\lambda\beta\cdots}
\right)=
\varepsilon^i{}_\mu\varepsilon^j{}_\nu\cdots
\varepsilon_k{}^\lambda\varepsilon_n{}^\beta\cdots
\frac{DT^{\mu\nu\cdots}_{\lambda\beta\cdots}}{dt}\ ,
\label{genembcon}
\end{equation}
where the covariant derivative reads
\begin{equation}
\frac{DT^{\mu\nu\cdots}_{\lambda\beta\cdots}}{dt}=
v^\lambda\left(
\partial_\lambda T^{\mu\nu\cdots}_{\alpha\beta\cdots}
+ \Gamma_{\bar{\mu}\lambda}{}^\mu T^{\bar{\mu}\nu\cdots}_{\alpha\beta\cdots}
+\cdots -\Gamma_{\alpha\lambda}{}^{\bar{\lambda}}
T^{{\mu}\nu\cdots}_{\bar{\lambda}\beta\cdots} -\cdots
\right)\ .
\label{covdir}
\end{equation}
Doing the differentiation in the left-hand side of (\ref{genembcon})
and applying relations (\ref{invmet}) we find
\begin{equation}
v^\lambda \Gamma_{\nu\lambda}{}^\mu = \left(
\varepsilon^\mu,\frac{d}{dt}\varepsilon_\nu\right) =
-\left(\varepsilon_\nu,\frac{d}{dt}\varepsilon^\mu\right)\ ,
\label{embgam}
\end{equation}
which should hold for any curve in $[q]$ (for any $v^\mu$). Thus,
we conclude that
\begin{equation}
\Gamma_{\mu\nu}{}^\lambda = g^{\lambda\alpha} (\varepsilon_\alpha,
\varepsilon_{\mu,\nu})\ .
\label{embgam2}
\end{equation}
Equation (\ref{embgam}) ensures that along any curve
$q^\mu(t)$, the fields
$\varepsilon_i{}^\mu(q(t))$ and
$\varepsilon^{i\mu}(q(t))$
are transported parallel, as expressed by the relations
$D\varepsilon_i{}^\mu(q(t))/dt=0,~D\varepsilon^{i\mu}(q(t))/dt=0$.
Applying the covariant derivative $D/dt$
to the metric (\ref{met}) we obtain from the chain rule of
differentiation
$Dg_{\mu \nu}(q(t))/dt=0$
for any curve in $[q]$,
which ensures the compatibility of the induced
connection with the induced metric.

Thus we have succeeded in determining metric and parallel
transport in the space $[q]$ by
an embedding of
all paths in $[q]$ into the space of all paths in the
 bigger euclidean space.
The embedding of the path space implies the embedding of the tangent
space, thus determining the induced metric on $[q]$. By imposing
the condition that the parallel transport law is compatible with
the embedding of the tangent space, the connection in the
space $[q]$ is uniquely determined, too.

\vskip 0.3cm
\noindent
{\bf 3}. Let us now turn to the analysis of the connection (\ref{embgam2}).
First of all, we observe that the torsion tensor is, in general, nonzero
\begin{equation}
S_{\nu\kappa}{}^\mu = \frac 12 g^{\mu\lambda}
\left[
(\varepsilon_\lambda,\varepsilon_{\nu,\kappa}) -(\varepsilon_\lambda,
\varepsilon_{\kappa,\nu})
\right]\ .
\label{embtor}
\end{equation}
The torsion induced by the embedding is zero iff
\begin{equation}
\varepsilon^i{}_{\nu,\mu} = \varepsilon^i{}_{\mu,\nu}\ .
\label{torzer}
\end{equation}
If this condition is satisfied, the matrix elements in
relations (\ref{tv}) are the derivatives
of $M$ functions $\varepsilon^i (q)$
\begin{equation}
\varepsilon^i{}_\mu(q) = \partial_\mu \varepsilon^i(q) \ .
\label{intcon}
\end{equation}
In this case, the path embedding (\ref{tv}) can be
achieved by a pointwise
embedding of the space $[q]$ in $[x]$. Relation
(\ref{tv}) can then be written in the form
\begin{equation}
dx^i = \varepsilon^i{}_\mu(q)dq^\mu = d \varepsilon^i(q)\ ,
\label{dx}
\end{equation}
i.e., we get the pointwise embedding
\begin{equation}
x^i = \varepsilon^i(q)\ .
\label{semb}
\end{equation}

The condition (\ref{tv}) may be thought as constraints
on the velocity $v^i$. The torsion tensor (\ref{embtor})
vanishes when the constraints are {\em integrable\/} as can be seen
from (\ref{torzer}). In this
case, the path embedding and the tangent space embedding can
be obtained right-away from the space embedding (\ref{semb}).
When the constraints are {\em non-integrable\/},
there is no pointwise embedding of $[q]$ into $[x]$, while
the path space or the tangent space can still be embedded
in the corresponding larger space.
The latter is sufficient to specify the metric and connection
induced by the embedding. In fact, in this approach the connection
appears to be the most general connection compatible with the metric.

The metric tensor $g_{\mu\nu}$ has $N(N+1)/2$ independent
components. The torsion tensor $S_{\nu\kappa}{}^\mu$ has
$N^2(N-1)/2$ independent components. To embed
a general metric space with torsion, the number $NM$, being
the number of independent
embedding coefficients $\varepsilon^i{}_\mu$, should be greater or equal
to $N(N^2+1)/2$. This leads to the relation between the dimensions
of the spaces $[q]$ and $[x]$
\begin{equation}
(\dim[q])^2 + 1 \leq 2\dim[x]\ .
\label{embdim}
\end{equation}

\vskip 0.3cm
\noindent
{\bf 4}. We are now ready to show that
the autoparallel curves are
specially favored geometric curves in the
space $[q]$ since our embedding procedure
maps them into the straight lines in the embedding space $[x]$.
A straight line parallel-transports its tangent vector
along itself with respect to a trivial connection $\Gamma_{ij}{}^k=0$:
\begin{equation}
\frac{Dv^i}{dt}=\dot{v}^i = 0\ .
\label{stline}
\end{equation}
Applying the compatibility condition
(\ref{embcon}) to the velocity vector $T^\mu=v^\mu$, we conclude
that the straight line is the image of a curve
satisfying the equation $Dv^\mu/dt =0$,
which is the autoparallel, as announced above.
Indeed, upon
substituting (\ref{tv}) into (\ref{stline}) and multiplying the result
by $\epsilon^{i\mu}$ we get
\begin{equation}
(\varepsilon^\mu, \dot{v})=
\dot{v}^\mu + \left(\epsilon^\mu, \frac{d}{dt} \epsilon_\nu
\right) \, v^\nu= \frac{Dv^\mu}{dt}= 0\ .
\label{auto2}
\end{equation}
For every path in $[q]$, the compatibility of the parallel transport
law with the embedding yields the condition (\ref{embgam}). Therefore
equation (\ref{auto2}) turns into the autoparallel equation (\ref{auto}).
In particular, when the constraint (\ref{tv}) is integrable, the
torsion vanishes, and equation (\ref{auto2}) describes geodesics
in the embedded manifold (\ref{semb}).

\vskip 0.3cm
\noindent
{\bf 5}. It is useful
to give a mechanical interpretation
of why autoparallels should be favored as
particle trajectories.
They describe a constrained motion with
an acceleration that deviates  {\em minimally\/} from the acceleration
of the corresponding unconstrained motion. This property can be
formulated mathematically by means of
Gauss' principle of least constraint \cite{arnold,var}.

Consider a Lagrangian system in the space $[x]$ with a Lagrangian
$L=L(x,v)$. At each moment of time,
a state of the system can labeled by the pair
$\psi = (x^i(t),v^i(t))$. At the state $\psi$,
construct a matrix $H_{ij} = \partial^2 L/\partial v^i\partial v^j$
called a Hessian of the system. Consider two paths $x^i_1(t)$ and
$x^i_2(t)$ going through the state $\psi$. Gauss' deviation
function (sometimes also called Gauss' constraint) for two paths
at the state $\psi$ reads
\begin{equation}
G_\psi = \frac 12 \left(\dot{v}^i_1 -\dot{v}^i_2\right)H_{ij}
 \left(\dot{v}^j_1 -\dot{v}^j_2\right)\ .
\label{gauss}
\end{equation}
It measures the deviation of two motions from one another at the same
system state \cite{arnold,var}.
Let the motion in $[x]$ be subject to constraints. All paths
$x^i(t)$ allowed by the constraints and going through a state $\psi$
are called {\em conceivable} motions. A path $\bar{x}^i(t)$ is called
{\em released} motion if it satisfies the Euler-Lagrange equations
for the Lagrangian $L$. Gauss' principle of least constraint says that
the deviation of conceivable motions from a released motion takes
a stationary value on the {\em actual} motion.

In our case, the released motion is a free motion with zero
acceleration $\ddot{\bar{x}}^i =0$ and $H_{ij} =\delta_{ij}$.
Accelerations of the conceivable motions are
\begin{equation}
\dot{v}^i = \varepsilon^i{}_\mu\dot{v}^\mu + \varepsilon^i{}_{\mu,\nu}
v^\mu v^\nu\ .
\label{acce}
\end{equation}
Gauss' deviation function (\ref{gauss}) assumes the form
\begin{equation}
G_\psi = \frac 12 \left[\dot{v}^\mu + \left(\varepsilon^\mu,
\frac{d}{dt}\varepsilon_\nu\right)v^\nu\right]^2\ .
\label{gauss2}
\end{equation}
It is non-negative, $G_\psi\geq 0$, for any state $\psi$ and
achieves its absolute minima, $G_\psi=0$, for the acceleration determined
by equation (\ref{auto2}).  Thus, the actual motion is realized by
autoparallels.

Gauss' principle of least constraint implies that the geometrical
force caused by the constraints,
must be {\em minimal\/} for the actual constrained motion.
Thus, in the framework of constrained dynamics,
autoparallels have the least deviation from the straight lines
describing free, unconstrained motions.
That is why they can rightfully
be called the {\em straightest\/} lines on a manifold.
The fact that particles must follow straightest lines is a
consequence of the physical phenomenon {\em inertia}. It is
hard to understand how a particle should know where to go
to make its trajectory the shortest path to a distant point.

Finally, we remark
that in the case of integrable constraints, Gauss' principle
of least constraint leads to geodesics, while for non-integrable
constraints, geodesics do not have the least deviation from
the free, unconstrained motion and do not play any special
role in the dynamics.

Another interpretation of the autoparallel equation (\ref{auto2})
rests on the d'Alembert-Lagrange principle \cite{arnold,var}.
In theoretical mechanics, elements of the tangent space
are also called {\em virtual velocities}. The embedding
condition (\ref{emb}) determines virtual velocities of the
constrained motion. Let us denote the Lagrange derivative
as $[L]_i =d/dt \partial L/\partial v^i - \partial L/\partial x^i$.
The d'Alembert-Lagrange principle
asserts that the conceivable motion of a system with the
Lagrangian $L$ is an actual motion if for every moment
of time
\begin{equation}
(T,[L]) = 0\ ,
\label{dlp}
\end{equation}
for all virtual velocities of the constrained motion.
Taking the free motion Lagrangian $L= (v,v)/2$ with the
constraint (\ref{tv}) and substituting them into (\ref{dlp})
we find that the autoparallel equation (\ref{auto2})
follows from (\ref{dlp}) for an arbitrary virtual velocity
$T^\mu$.

In addition we remark that autoparallels can be obtained
from H\"older's variational principle \cite{arnold,var}
applied to the free action. Let $\delta x^i \in T[x]$
be a variation vector field. Amongst all variation
vector fields we single out those that are virtual
velocities of the constrained motion,
\begin{equation}
\delta x^i
= \varepsilon^i{}_\mu \delta q^\mu\ ,\
\ \ \ \delta q^\mu \in T[q]\ .
\label{vvf}
\end{equation}
A conceivable path is called a critical point of the action
functional if its variation vanishes when restricted on
the subspace of virtual velocities of the constrained motion.
H\"older's variational principle suggests that the actual
constrained motion is a critical point of the action.
Making a variation of the action
we find the equation of motion
\begin{equation}
(\delta x^i,[L])=0\ .
\label{xl}
\end{equation}
Substituting the expression (\ref{tv}) for conceivable velocities
and admissible variations (\ref{vvf}) in (\ref{xl}), we obtain
the autoparallel equation (\ref{auto2}) for the free Lagrangian,
$[L]_i = \delta_{ij}\dot{v}^j$.

Another remark concerns relativistic motion.
The autoparallel motion may also be embedded into
a Minkovski space in the same fashion as for a euclidean
space. In all formulas given above, the euclidean metric
$\delta_{ij}$ has to be replaced by a corresponding indefinite
metric of the Minkovski space. Similarly, in the three variational
principles for autoparallels considered in this section, the free
motion in the embedding space should be described by the
corresponding Lagrangian of a free relativistic particle,
while all time derivatives in the equations of motion
must be replaced by the derivatives with
respect to the proper time defined on the constraint surface.

Finally we point out that the motion of a {\em holonomic}
system is completely determined by the restriction
of the Lagrangian to the constraint surface \cite{arnold}.
Thus, holonomic constrained systems are indistinguishable
from ordinary unconstrained Lagrangian systems. This
is not true for non-holonomic systems, meaning that the
Euler-Lagrange equations for the Lagrangian restricted
on the constraint surface do not coincide with the original
equations for the constrained motion.
This difficulty
prevents us from applying a conventional Hamiltonian
formalism to the autoparallel motion, and subjecting it to
a  canonical quantization. In other words, Dirac's method
of quantizing constrained systems \cite{dirac} does not
apply to non-holonomic systems because their motion is
not described by the conventional Lagrange formalism
\cite{arnold}.
The problem requires a further study.

\vskip 0.3cm
\noindent
{\bf Acknowledgment}

\vskip 0.2cm
Drs. G. Barnich and A. Tchervyakov are thanked for useful discussions.
This work was partially supported by the Alexander von
Humboldt foundation.

\end{document}